\documentclass[submission,copyright,creativecommons]{eptcs}
\providecommand{\event}{FOCLASA 2012} 

\usepackage{graphicx}
\usepackage{amsmath}
\usepackage{amsfonts}
\usepackage{amssymb}
\usepackage{hyperref}
\usepackage{url}
\usepackage{color}
\usepackage{rotating}
\usepackage[caption=false,font=footnotesize]{subfig}

\newcommand{\goes}[1]{\xrightarrow[]{#1}}

\newtheorem{prop}{Proposition}
\newtheorem{defn}{\ \\Definition}

\title{A multi-level model for self-adaptive systems}
\author{Emanuela Merelli \qquad\qquad Nicola Paoletti \qquad\qquad Luca Tesei
\institute{School of Science and Technology\\Computer Science Division, University of Camerino\\Camerino, IT}
\email{\quad emanuela.merelli@unicam.it \quad\qquad nicola.paoletti@unicam.it \quad\qquad luca.tesei@unicam.it}
}
\def\titlerunning{A multi-level model for self-adaptive systems}
\def\authorrunning{E. Merelli, N. Paoletti \& L. Tesei}
\begin{document}
\maketitle

\begin{abstract}
This work introduces a general multi-level model for self-adaptive systems. A
self-adaptive system is seen as composed by two levels: the lower level
describing the actual behaviour of the system and the upper level accounting for
the dynamically changing environmental constraints on the system. In order to
keep our description as general as possible, the lower level is modelled as a
state machine and the upper level as a
second-order state machine whose states have associated formulas over observable variables of the lower level.
Thus, each state of the second-order machine identifies the set of lower-level
states satisfying the constraints. Adaptation is triggered when a second-order
transition is performed; this means that the current system no longer can
satisfy the current high-level constraints and, thus, it has to adapt its
behaviour by reaching a state that meets the new constraints. The semantics of
the multi-level system is given by a flattened transition system that can be
statically checked in order to prove the correctness of the adaptation model. To
this aim we formalize two concepts of weak and strong adaptability providing
both a relational and a logical characterization. We report that this work gives
a formal computational characterization of multi-level self-adaptive systems,
evidencing the important role that (theoretical) computer science could play in
the emerging science of complex systems.
\end{abstract}

\section{Introduction}
Self-adaptive systems are a particular kind of systems able to modify their own behaviour according to their environment and to their current configuration. They learn from the environment and develop new strategies in order to fulfil an objective, to better respond to problems, or more generally to maintain desired conditions. Self-adaptiveness is an intrinsic property of the living matter. Complex biological systems naturally exhibit auto-regulative mechanisms that continuously trigger internal changes according to external stimuli. Moreover, self-adaptation drives both the evolution and the development of living organisms.

Recently there has been an increasing interest in self-adaptive properties of software systems. In~\cite{laddaga} the following definition is given: \textit{``Self-adaptive software evaluates its own behaviour and changes behaviour when the evaluation indicates that it is not accomplishing what the software is intended to do, or when better functionality or performance is possible.''}

As a matter of fact, software systems are increasingly resembling complex systems and they need to dynamically adapt in response to changes in their operational environment and in their requirements/goals. Two different types of adaptation are typically distinguished:
\begin{itemize}
\item \textit{Structural adaptation}, which is related to architectural reconfiguration. Examples are addition, migration and removal of components, as well as reconfiguration of interaction and communication patterns.
\item \textit{Behavioural adaptation}, which is related to functional changes, e.g.\ changing the program code or following different trajectories in the state space.
\end{itemize}

Several efforts have been made in the formal modelling of self-adaptive
software, with particular focus on verifying the correctness of the system
after adaptation. Zhang et al.\ give a general state-based model of
self-adaptive programs, where the adaptation process is seen as a transition
between different non-adaptive regions in the state space of the
program~\cite{zhang2006model}. In order to verify the correctness of adaptation
they define a new logic called A-LTL (an adapt-operator extension to LTL) and
model-checking algorithms~\cite{zhang2009modular} for verifying adaptation
requirements. In PobSAM~\cite{khakpour2010pobsam,khakpour2010formal} actors
expressed in Rebeca are governed by managers that enforce dynamic policies
(described in an algebraic language) according to which actors adapt their
behaviour. Different adaptation modes allow to handle events occurring during
adaptation and ensuring that managers switch to a new configuration only once
the system reaches a safe state. Another example is the work by Bruni et
al.~\cite{bruni2012conceptual} where adaptation is defined as the run-time
modification of the control data and the approach is instantiated into a formal
model based on labelled transition systems. In~\cite{bruni2008modelling},
graph-rewriting techniques~\cite{le1998describing} are employed to describe
different characterizations of dynamical software architectures. Meseguer and
Talcott~\cite{meseguer2006semantic} characterize adaptation in a model for
distributed object reflection based on rewriting logic and nesting of
configurations. Theorem-proving techniques have also been used for assessing the
correctness of adaptation: in~\cite{kulkarni2004correctness} a proof lattice
called transitional invariant lattice is built to verify that an adaptive
program satisfies global invariants before and after adaptation. In particular
it is proved that if it is possible to build that lattice, then adaptation is
correct.

There are several other works worth mentioning, but here we do not aim at
presenting an exhaustive state-of-the-art in this widening research field. We
address the interested reader to the
surveys~\cite{cheng2009software,salehie2009self} for a general introduction to
the essential aspects and challenges in the modelling of self-adaptive software
systems.

\subsection{A multi-level view of self-adaptation}
Complex systems can be regarded as multi-level systems, where two fundamental
levels can be distinguished: a \textbf{behavioural level} $B$ accounting for the
dynamical behaviour of the system; and a higher \textbf{structural level} $S$
accounting for the global and more persistent features of the system. These two
levels affect each other in two directions: \textit{bottom-up}, e.g.\ when a
collective global behaviour or new emergent patterns are observed; and
\textit{top-down}, e.g.\ when constraints, rules and policies are superimposed
on the behavioural level. These two fundamental levels and their relationships
are the base to scale-up to multi-level models. In a generic multi-level model,
any $n$-th level must resemble the behavioural level, the corresponding
$n+1$-level has to match with the structural level and the relationships between
them will have to show the same characteristics. We discuss how this scale-up is
implemented in our setting in~Section~\ref{sect:conclusion}

Multiple levels arise also when software systems are concerned. For instance,
in~\cite{corradini2006relating} Corradini et al. identify and formally relate
three different levels: the \textit{requirement level}, dealing with high-level
properties and goals; the \textit{architectural level}, focusing on the
component structure and interactions between components; and the
\textit{functional level}, accounting for the behaviour of a single component.
Furthermore, Kramer and Magee~\cite{kramer2007self} define a three-level
architecture for self-managed systems consisting of a \textit{component control
level} that implements the functional behaviour of the system by means of
interconnected components; a \textit{change management level} responsible for
changing the lower component architecture according to the current status and
objectives; and a \textit{goal management level} that modifies the lower change
management plans according to high-level goals. Hierarchical finite state
machines and Statecharts~\cite{harel1987statecharts} have also been employed to
describe the multiple architectural levels in self-adaptive software
systems~\cite{karsai2003approach,shin2005self}.

In this work we introduce $S[B]$\textit{-systems}: a general state-based model
for self-adaptive systems where the lower behavioural level describes the actual
dynamic behaviour of the system and the upper structural level accounts for the
dynamically changing environmental constraints imposed on the lower system. The
$B$-level is modelled as a state machine $B$. The upper
level is also described as a state machine where each state
has associated a set of constraints (logical formulas) over variables resulting
from the observation of the lower-level states, so that each $S$-state
identifies the set of $B$-states satisfying
the constraints. Therefore, a set of dynamically changing constraints underlies
a \textit{second-order structure} $S$ whose states are sets of $B$-states and,
consequently, transitions relate sets of $B$-states.

We focus on \textit{behavioural and top-down adaptation}: the $B$-level adapts
itself according to the higher-level rules. In other words the upper level
affects and constrains the lower level. Adaptation is expressed by firing a
higher-order transition, meaning that the $S$-level switches to a different set
of constraints and the $B$-level has adapted its behaviour by reaching a state
that meets the new constraints. Our idea is broadly inspired by Zhang et
al.~\cite{zhang2006model}, i.e.\ the state space of an adaptive program can be
separated into a number of regions exhibiting a different \textit{steady-state
behaviour} (behaviour without reconfiguration). However, in our model the
steady-state regions are represented in a more declarative way using
constraints associated to the states of the $S$-level. Moreover, in
$S[B]$-systems not only the behavioural level, but also the adaptation model
embedded in the structural level is dynamic. Adaptation of the $B$-level is not
necessarily instantaneous and during this phase the system is left unconstrained
but an invariant condition that is required to be met during adaptation.
Differently to~\cite{zhang2006model}, the invariants are specific for every
adaptation transition making this process controllable in a finer way. The
semantics of the multi-level system is given by a flattened transition system
that can be statically checked in order to prove the correctness of the
adaptation model. To this aim we also formalize the notion of adaptability,
i.e.\ the ability of the behavioural level to adapt to a given structural level.
We distinguish between weak and strong adaptability, providing both a relational
and a logical characterization for each of them.

$S[B]$-systems has been inspired by some of the authors' recent work in the
definition of a spatial bio-inspired process algebra called \textit{Shape
Calculus}~\cite{shape1,shape2}. In that case, a process $S[B]$ is characterized
by a reactive behaviour $B$ and by a shape $S$ that imposes a set of geometrical
constraints on the interactions and on the occupancy of the process. This idea
is shifted in a more general context in the $S[B]$-systems where, instead, we
consider sets of structural constraints on the state space of the $B$-level.
We want to underline that previous work and, mainly, this work have been
conceived as contributions not only in the area of adaptive software system, but
also in the area of modelling complex natural systems.

The notion of multiple levels that characterizes our approach for computational
adaptive systems is something well-established in the science of complex
systems. As pointed out by Baianu and Poli~\cite{baianu2010simple} \textit{``All
adaptive systems seem to require at least two layers of organization: the first
layer of the rules governing the interactions of the system with its environment
and with other systems, and a higher-order layer that can change such rules of
interaction.''} $S[B]$-systems are similarly built on two levels: the $B$-level
describes the state-based behaviour of the system and the $S$-level regulates
the dynamics of the lower level. In our settings, communication and interactions
are not explicitly taken into account. Indeed the behavioural finite state
machine can describe the semantics of a system made by several interacting
components.

Another accepted fact is that higher levels in complex adaptive systems lead to
higher-order structures. Here the higher $S$-level is described by means of a
second order state machine (i.e.\ a state machine over the powerset of the
$B$-states). Similar notions have been formalized by Baas~\cite{nils3baas} with
the \textit{hyperstructures} framework for multi-level and higher-order
dynamical systems; and by Ehresmann and Vanbremeersch with their \textit{memory
evolutive systems}~\cite{ehresmann2007memory}, a model for hierarchical
autonomous systems based on category theory.

The paper is organized as follows. Section~\ref{sect:model} introduces the
formalism and the syntax of $S[B]$-systems, together with an ecological example
that will be used also in the following. In Section~\ref{sect:semantics} we give
the operational semantics of a $S[B]$-system by means of a flattened transition
system. In Section~\ref{sect:adaptability} we formalize the concepts of weak and
strong adaptability both in a relational and in a logical form. Finally,
conclusions and possible future developments of the model are discussed in
Section~\ref{sect:conclusion}.

\section{A multi-level state-based model}\label{sect:model}
An $S[B]$-system encapsulates both the behavioural ($B$) and the
structural/adaptive ($S$) aspects of a system. The behavioural level is
classically described as a finite state machine of the form $B=(Q, q_0,
\goes{}_B)$. In the following, the states $q \in Q$ will also be referred to as
$B$-states and the transitions as $B$-transitions.

The structural level is modelled as a finite state machine $S =
(R, r_0, \goes{}_S, L)$ ($R$ set of states, $r_0$ initial state, $\goes{}_S$
transition relation and $L$ state labelling function). In the following, the
states $r \in R$ will be also referred to as $S$-states and the transitions as
$S$-transitions. The function $L$ labels each $S$-state with a set of formulas
(the constraints) over an \emph{observation} of the $B$-states in the form of a
set of variables $X$. Therefore an $S$-state $r$ uniquely identifies the set of
$B$-states satisfying $L(r)$ and $S$ gives rise to a second-order structure $(R
\subseteq 2^Q, r_0, \goes{}_S \subseteq 2^Q \times 2^Q, L)$.

In this way, behavioural adaptation is achieved by switching from an $S$-state
imposing a set of constraints to another $S$-state where a (possibly) different
set of constraints holds. During adaptation the behavioural level is no more
regulated by the structural level, except for a condition, called
\textit{transition invariant}, that must be fulfilled by the system undergoing
adaptation. We can think of this condition as a minimum requirement to which the
system must comply to when it is adapting and, thus, it is not
constrained by any $S$-state.

Note that an $S[B]$-system \textit{dynamically} adapts and reconfigures its
behaviour, thus both the behavioural level and the structural level are dynamic.

\begin{defn}[$S\text{[}B\text{]}$-system behaviour]
The behaviour of an $S[B]$-system $S[B]$ is a tuple $B=(Q, q_0, \goes{}_B)$, where
\begin{itemize}
\item $Q$ is a finite set of states and $q_0 \in Q$ is the initial state; and
\item $\goes{}_B \subseteq Q \times Q$ is the transition relation.
\end{itemize}
\end{defn}

In general, we assume no reciprocal internal knowledge between the $S$- and the
$B$-level. In other words, they see each other as \textit{black-box systems}.
However, in order to realize our
notion of adaptiveness, there must be some information flowing bottom-up from
$B$ to $S$ and some information flowing top-down from $S$ to $B$. In particular,
the bottom-up flow is modelled here as a set of variables $X = \{x_1, \ldots,
x_n\}$
called \emph{observables} of the $S$-level on the $B$-level.
The values of these variables must always be \emph{derivable} from the
information contained in the $B$-states, which can possibly hold more ``hidden''
information related to internal activity. This keeps our approach
black-box-oriented because the $S$-level has not the full knowledge of the
$B$-level, but only some derived (e.g.\ aggregated, selected or calculated)
information.
Concerning the top-down flow, the $B$-system only knows whether its current
state satisfies the current constraint or not. If not, we can assume that the
possible target $S$-states and the relative invariants are outputted by the
$S$-system and given in input to the $B$-system.


\begin{defn}[$S\text{[}B\text{]}$-system structure]
The structure of an $S[B]$-system $S[B]$ is a tuple $S=(R, r_0, \goes{}_S, L)$,
where
\begin{itemize}
\item $R$ is a finite set of states and $r_0 \in R$ is the initial state;
\item $\goes{}_S \subseteq R \times \Phi(X) \times R$ is a transition relation,
labelled with a formula called \textit{invariant}; and
\item $L:R\goes{}\Phi(X)$ is a function labelling each state with a formula
over a set of \emph{observables} $X=\{x_1, \ldots, x_n\}$.
\end{itemize}
\end{defn}

Thus, an $S[B]$-system has associated a finite set $X=\{x_1, \ldots, x_n\}$ of
typed variables over finite domains $\{D_1,\ldots, D_n\}$ whose values must be completely determined in
each state of $Q$. More formally,

\begin{defn}[Observation Function]
Given an $S[B]$-system $S[B]$ with a set $X=\{x_1, \ldots, x_n\}$ of
observables, an \emph{observation function} $\mathcal{O}\colon Q \rightarrow
\prod_{i=1}^n D_i$ is a total function that maps each $B$-state $q$ to the
tuple of variable values $(v_1, \ldots, v_n) \in D_1 \times \ldots \times D_n$
observed at $q$.
\end{defn}

Note that we do not require this function to be bijective. This means that some
different states can give the same values to the observables. In this case, the
difference is not visible to $S$, but it is internal to $B$.

We indicate with $\Phi(X)$ the set of formulas over the variables in $X$. We
assume that constraints are specified with a first-order logic-like language.

\begin{defn}[Satisfaction relation]
Let $S[B]$ be a $S[B]$-system with a set $X=\{x_1, \ldots, x_n\}$ of observables
and with an observation function $\mathcal{O}$. A state $q \in Q$
\emph{satisfies} a formula $\varphi \in \Phi(X)$, written $q \models \varphi$,
iff $\varphi$ is satisfied applying the substitution $\{^{v_1}/x_1, \ldots,
^{v_n}/x_n\}$, where $\mathcal{O}(q) = (v_1, \ldots, v_n)$, using the
interpretation rules of the logic language.

Let us also define an evaluation function $[[\_]]:
\Phi(X) \goes{} 2^Q$ mapping a formula
$\varphi \in \Phi(X)$ to the set of $B$-states $Q' = \lbrace q \in Q \ | \ q
\models \varphi \rbrace$, i.e.\ those satisfying $\varphi$.
\end{defn}

%

Let us now give an intuition of the adaptation semantics. Let the active
$S$-state be $r_i$ and $r_i \goes{\varphi}_S r_j$. Assume that the behaviour is
in a steady state (i.e.\ not adapting) $q_i$ and therefore $q_i \models L(r_i)$.
If there are no $B$-transitions $q_i \goes{}_B q_j$ such that $q_j \models
L(r_i)$
the system starts adapting to the target $S$-state $r_j$. In this phase, the
$B$-level is no more constrained, but during adaptation the invariant $\varphi$
must be met. Adaptation ends when the behaviour reaches a state $q_k$ such that
$q_k \models L(r_j)$.

The following definition determines when the structure $S$ of a $S[B]$-system
is well formed, that is: it must no contain inconsistencies w.r.t.\ all possible
variable observations and the initial $B$-state must satisfy the initial
$S$-state.

\begin{defn}[Well-formed structure] Let $S[B]$ be a $S[B]$-system. The
structural level $S$ is well-formed if the following conditions hold:
\begin{itemize}
\item for all $S$-states $r \in R$, $L(r)$ must be \textbf{satisfiable}, in the
sense that there must be a variable observation under which $L(r)$ holds
($\exists q \in Q. \ q \models L(r)$) and
\item the initial $B$-state must satisfy the constraints in the initial
$S$-state, i.e.\ $q_0 \models L(r_0)$.
\end{itemize}
\end{defn}

In the remainder of the paper we assume to deal with well-formed structures
without explicitly mentioning it.

\subsection{An example from ecology}\label{subsect:exmpl}
\begin{figure}
\centering
\includegraphics[width=\textwidth]{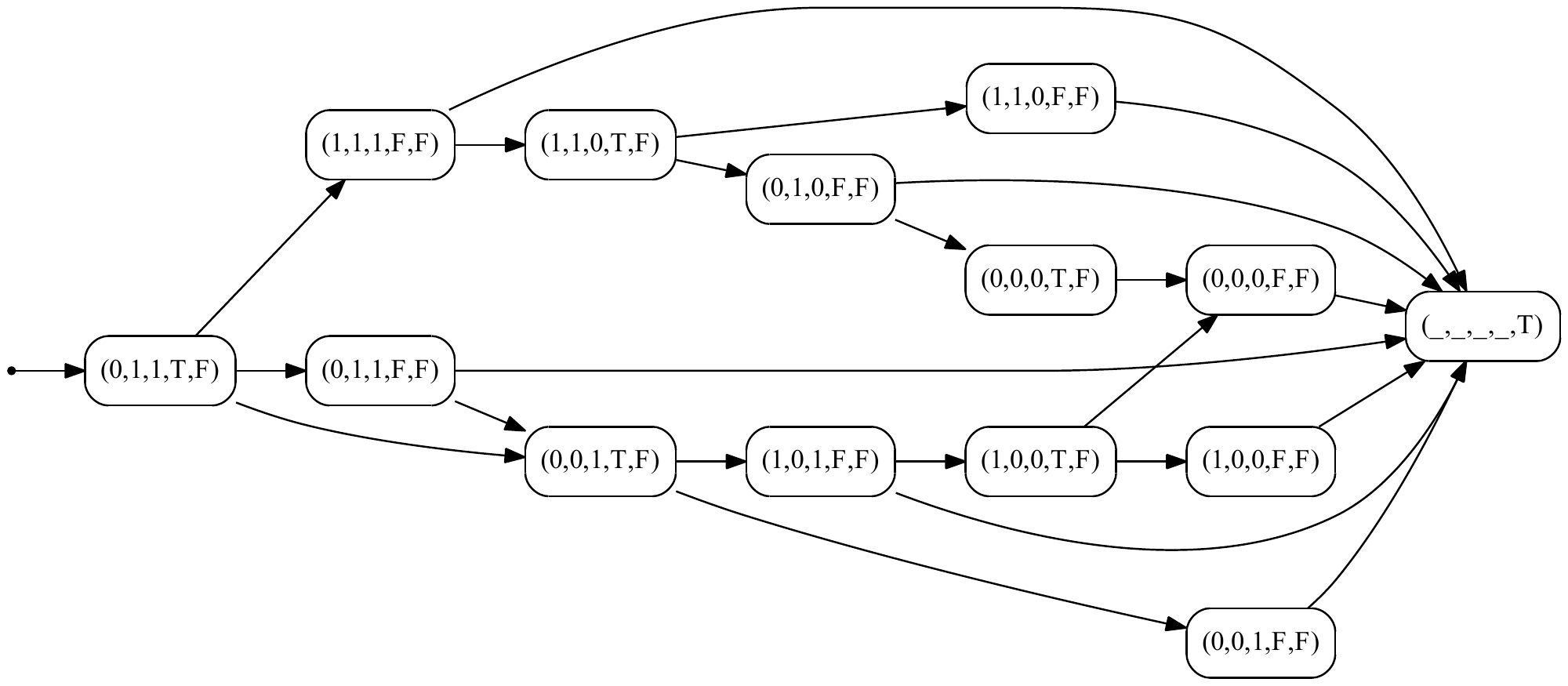}
\caption{The behavioural state machine $B$ for the adaptive 1-predator 2-prey food web example. Each state is characterized by a different combination of the variables $(p, a_0, a_1, eat, moved)$ (favourite prey, availability of $p_0$, availability of $p_1$, has the predator eaten?, has the predator migrated?). The initial state is $(0,1,1,\mathsf{true}, \mathsf{false})$. All the states where $moved = \mathsf{true}$ has been grouped for simplicity to a single state $(\_,\_,\_,\_\mathsf{true})$.}\label{fig:b-level}
\end{figure}

In this part we introduce a case study in the field of ecology and population biology: \textit{the adaptive 1-predator 2-prey food web}. This system describes a variant of classical prey-predator dynamics where in normal conditions the predator consumes its favourite prey $p_0$. When the availability of $p_0$ is no longer sufficient for the survival of the predator, it has to adapt its diet to survive and it consequently starts consuming another species $p_1$. For the sake of showing the features of our model, here we present an oversimplified version of this system that omits quantitative aspects like predation rates and growth of prey. We assume that the predator initially consumes the prey $p_0$ (variable $p=0$) and that prey may be available (variable $a_i=1, \ i=0,1$) or not (variable $a_i=0, \ i=0,1$). The effect of consuming an available prey is to make that prey unavailable, as expected. The predator may also decide not to eat and change its diet (variable $p=1$). A boolean variable tells whether in the current state the predator has eaten some prey (variable $eat$). At each step the predator can do one of the following:
\begin{itemize}
\item eat the currently favourite prey $p_i$, if available ($a_i \leftarrow a_i - 1$ and $eat \leftarrow \mathsf{true}$);
\item do not eat and switch its favourite prey ($p \leftarrow |1 - p|$ and $eat \leftarrow \mathsf{false}$); or
\item do not eat.
\end{itemize}
Finally, if the predator does not feed itself for two consecutive times, it migrates to a more suitable habitat  (variable $moved = \mathsf{true}$) and no further actions are possible. The attentive reader may notice that under these restrictions the system will inevitably lead to a state where the predator moves to a different habitat. This is due to the fact that prey growth is not modelled here and it is always the case that the system eventually reaches a state where the predator cannot feed because of the unavailability of both prey. Each state of the behavioural level (depicted in Fig.~\ref{fig:b-level}) is described by a different evaluation of the involved variables:
$$(p, a_0, a_1, eat, moved) \in \lbrace 0, 1\rbrace \times \lbrace 0, 1\rbrace \times \lbrace 0, 1\rbrace \times \lbrace \mathsf{false}, \mathsf{true}\rbrace \times \lbrace \mathsf{false}, \mathsf{true}\rbrace.$$

\begin{figure}
\centering
\includegraphics[width=0.8\textwidth]{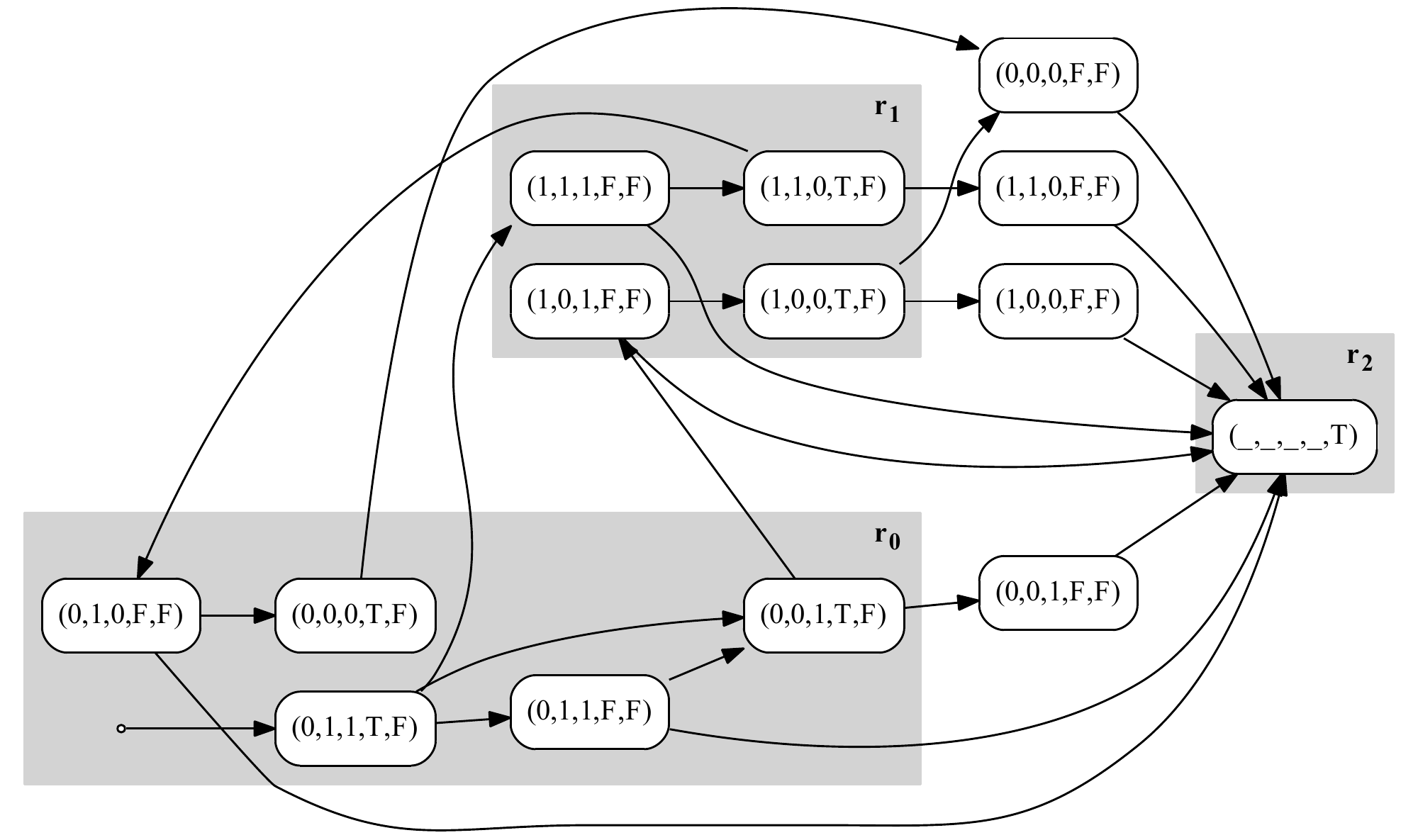}
\caption{$S$-states determining stable regions in the adaptive 1-predator 2-prey system.}\label{fig:b-level1}
\end{figure}

\begin{figure}
\centering
\subfloat[The state machine $S_0$]{\includegraphics[width=0.48\textwidth]{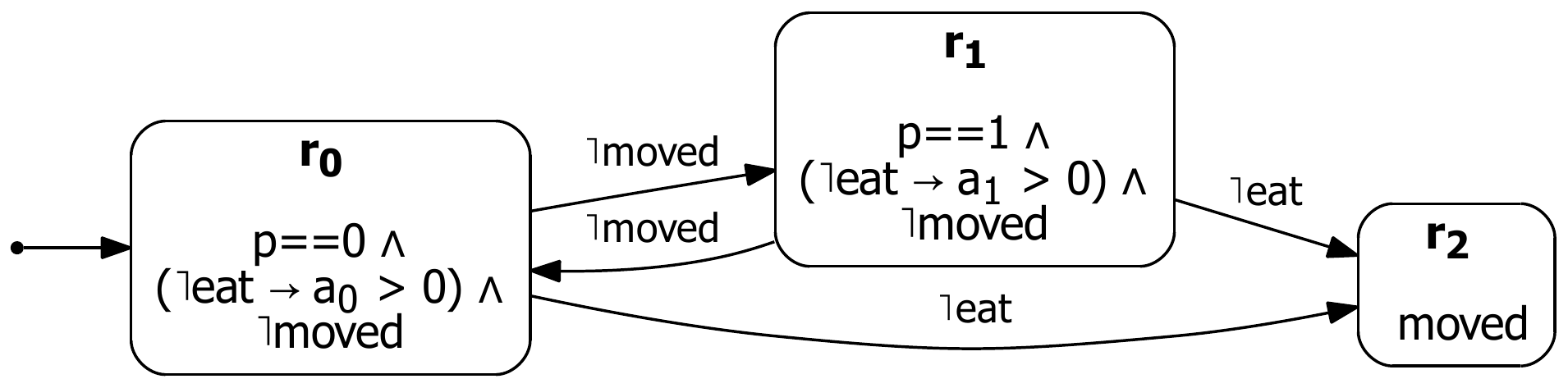}}\hfill \subfloat[The state machine $S_1$]{\includegraphics[width=0.48\textwidth]{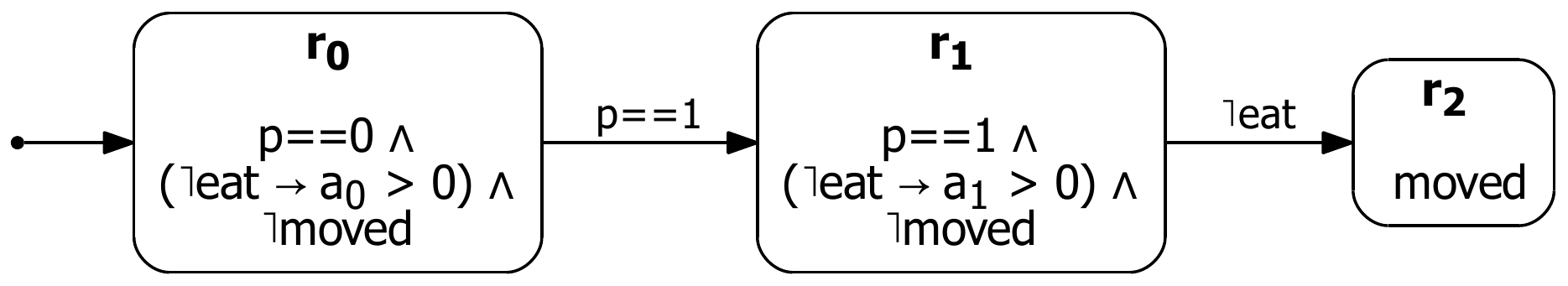}}
\caption{The two different structural levels $S_0$ and $S_1$ in the adaptive 1-predator 2-prey food web example. In each $S$-state $r_i$ the constraint imposed to the behavioural level are represented. Transition labels indicate adaptation invariants. $S_0$ allows the predator to adapt its diet and migrate due to starvation anytime. In $S_1$ adaptation is guided from $r_0$ (consume prey $p_0$), to $r_1$ (consume prey $p_1$) and finally to $r_2$ (migration).}\label{fig:s-level}
\end{figure}

In this example we consider two different $S$-levels (represented in Fig.~\ref{fig:s-level}): $S_0$ and $S_1$, but with the same set of $S$-states. More specifically $S_0$ is given by:
\begin{align*}
R = \ & \ \lbrace r_0, r_1, r_2 \rbrace,\\
 \goes{}_S = \ & \ \lbrace r_0 \goes{\neg moved} r_1, r_0 \goes{\neg eat} r_2, r_1 \goes{\neg moved} r_0, r_1 \goes{\neg eat} r_2\rbrace\\
 L(r) = \ & \ \lbrace p==0 \wedge (\neg eat \implies a_0 > 0) \wedge \neg moved
\rbrace \ \text{ if } r=r_0\\
 \ & \ \lbrace p==1 \wedge (\neg eat \implies a_1 > 0) \wedge \neg moved \rbrace
\ \text{ if } r=r_1\\
 \ & \ \lbrace moved \rbrace \ \text{ if } r=r_2.
\end{align*}
On the other hand, $S_1$ differs from $S_0$ only in the transition function, that is:
$$\goes{}_S \ = \ \lbrace r_0 \goes{p==1} r_1, r_1 \goes{\neg eat} r_2\rbrace$$

The three different $S$-states model three different stable regions in the prey-predator dynamics:
\begin{itemize}
\item $r_0:$ the predator consumes $p_0$. More precisely, the constraints require that the favourite prey must be $p_0$ ($p==0$); that the predator has not moved to another habitat ($\neg moved$); and that if the predator is not currently feeding, the prey $p_0$ must be available so that the predator can eat in the following step ($\neg eat \implies a_0 > 0$).
\item $r_1:$ the predator consumes $p_1$; the constraints are the same as $r_0$, but referred to prey $p_1$.
\item $r_2:$ the predator has migrated.
\end{itemize}

Figure~\ref{fig:b-level1} shows how the structural constraints identify
different stable regions in the behavioural level. The adaptation dynamics,
regulated by the transitions in $S_0$, allow the predator to adapt from $r_0$ to
$r_1$, under the invariant $\neg moved$ indicating that during adaptation the
predator cannot migrate. The equivalent $S$-transition is defined from $r_1$ to
$r_0$, so that the predator is able to return to its initially favourite prey.
Both from $r_0$ and $r_1$ a $S$-transition to $r_2$ is allowed under the
invariant $\neg eat$. In this way, the predator can adapt itself and migrate to
a different habitat under starvation conditions. On the other hand, the
transition relation in $S_1$ has been defined in a simpler way, which makes the
predator adapt deterministically from $r_0$ to $r_1$ and finally to $r_2$. In
this case, the adaptation invariant from $r_0$ to $r_1$ requires that the
predator has changed its diet to prey $p_1$.

The following section will show the operational rule for deriving the
transitional semantics of the $S[B]$-system as a whole and the semantics of
$S_0[B]$ and $S_1[B]$ in the adaptive 1-predator 2-prey system will be given as
well.

\def\name#1{\mbox{\sc #1}}

\newcommand{\equaldef}{\stackrel{\mbox{\tiny{def}}}{=}}
\def\sos#1#2{{\def\arraystretch{1.6}\begin{array}{c}#1\\\hline
#2\end{array}}}
\newcommand{\nar}[1]{\xrightarrow{#1}}

\section{Operational semantics}\label{sect:semantics}
In this part, we give the operational semantics of an $S[B]$-system as a transition system resulting from the flattening of the behavioural and of the structural levels. We obtain a Labelled Transition System (LTS) over states of the form $(q, r, \rho)$, where
\begin{itemize}
\item $q \in Q$ and $r \in R$ are the active $B$-state and $S$-state, respectively; and
\item $\rho$ keeps the target $S$-state that can be reached during adaptation and the invariant that must be fulfilled during this phase. Therefore $\rho$ is either empty (no adaptation is occurring), or a singleton $\lbrace (\varphi, r') \rbrace$, with $\varphi \in \Phi(X)$ a formula and $r' \in R$ an $S$-state.
\end{itemize}
\begin{defn}[Flat $S\text{[}B\text{]}-system$]
Let $S[B]$ be an $S[B]$-system. A flat $S[B]$-system is a LTS
$F(S[B])=(F,f_0,\goes{r} \cup \goes{r,\varphi,r'})$ where
\begin{itemize}
\item $F \subseteq Q \times R \times 2^{\Phi(X) \times R}$ is the set of
states;
\item $f_0 = (q_0, r_0, \emptyset)$ is the initial state;
\item $\goes{r} \subseteq F \times F$, with $r \in R$, is a family of transition relations between non-adapting states satisfying $L(r)$; and
\item $\goes{r,\varphi,r'} \subseteq F \times F$, with $r,r' \in R$ and $\varphi \in \Phi(X)$, is a family of transition relations between states during the adaptation determined by the $S$-transition $r \goes{\varphi}_S r'$. As a consequence it holds that for all $r,r',\varphi$, $\goes{r} \cap \goes{r,\varphi,r'} = \emptyset$.
\end{itemize}
\end{defn}

\begin{table}
\begin{small}
\[
\begin{array}{|c|}
\hline
\name{Steady}\sos{q\goes{}_Bq' \quad q' \models L(r)}{(q, r, \emptyset) \goes{r} (q', r, \emptyset)}
\\
\name{AdaptStart}\sos{\forall q''.(q \goes{}_Bq'' \implies q'' \not\models L(r)) \quad q\goes{}_Bq' \quad r\goes{\varphi}_Sr' \quad q' \models \varphi}{( q, r, \emptyset) \goes{r,\varphi,r'} ( q', r, \lbrace(\varphi,r')\rbrace)}
\\
\name{Adapt}\sos{q\goes{}_Bq' \quad q' \models \varphi \quad q \not\models L(r')}{( q, r, \lbrace(\varphi,r')\rbrace) \goes{r,\varphi,r'} ( q', r, \lbrace(\varphi,r')\rbrace)}
\\
\name{AdaptEnd}\sos{q \models L(r')}{(q, r, \lbrace(\varphi,r')\rbrace) \goes{r,\varphi,r'} ( q, r', \emptyset)}
\\
\hline
\end{array}\]
\end{small}
\caption{Operational semantics of a $S[B]$-system}\label{tbl:sos}
\end{table}

Table~\ref{tbl:sos} lists the set of rules characterizing the flattened transitional semantics of an $S[B]$-system:
\begin{itemize}
\item Rule \textsc{Steady} describes the steady (i.e.\ non-adapting) behaviour
of the system. If the system is not adapting and the $B$-state $q$ can perform a
transition to a $q'$ that satisfies the current constraints $L(r)$, then the
flat system can perform a non-adapting transition $\goes{r}$ of the form $(q, r,
\emptyset) \goes{r} (q', r, \emptyset)$.
\item Rule \textsc{AdaptStart} regulates the starting of an adaptation phase. Adaptation occurs when none of the next $B$-states satisfy the current specification ($\forall q''.(q \goes{}_Bq'' \implies q'' \not\models L(r))$, or more compactly $(q, r, \emptyset) \not\goes{r}$). In this case, for each $S$-transition $r \goes{\varphi}_S r'$ an adaptation towards the target state $r'$ under the invariant $\varphi$ starts and the flat system performs an adapting transition $\goes{r,\varphi,r'}$ of the form $(q, r, \emptyset) \goes{r,\varphi,r'} (q', r, \lbrace(\varphi,r')\rbrace)$.
\item Rule \textsc{Adapt} describes the evolution during the actual adaptation,
leading to transitions of the form $(q, r, \lbrace(\varphi,r')\rbrace)
\goes{r,\varphi,r'} (q', r, \lbrace(\varphi,r')\rbrace)$. During adaptation the
behaviour is not regulated by the specification and it must not satisfy the
target constraints $L(r')$ ($q \not\models L(r')$). We also require that the
invariant $\varphi \in \Phi(X)$ must always hold during this phase. Note that
the semantics does not immediately assure that a state where the target formula
holds is eventually reached. Formulations of the adaptability requirement are
given in Section~\ref{sect:adaptability}.
\item Rule \textsc{AdaptEnd} describes the end of the adaptation phase, i.e.\ a transition $\goes{r,\varphi,r'}$ from an adapting state $(q, r, \lbrace(\varphi,r')\rbrace)$ where $q$ satisfies the set of target constraints ($q' \models L(r')$), to the steady non-adapting state $(q, r', \emptyset)$.
\end{itemize}
Note that rules \textsc{Steady+AdaptStart} ensure that there cannot exist a non-adapting state with both an outgoing non-adapting transition $\goes{r}$ and an outgoing adapting transition $\goes{r,\varphi,r'}$. Conversely, rules \textsc{Adapt+AdaptEnd} ensure that there cannot exist an adapting state with both an outgoing non-adapting transition and an adapting transition.

\begin{sidewaysfigure}
\centering
\subfloat[]{\includegraphics[scale=0.4]{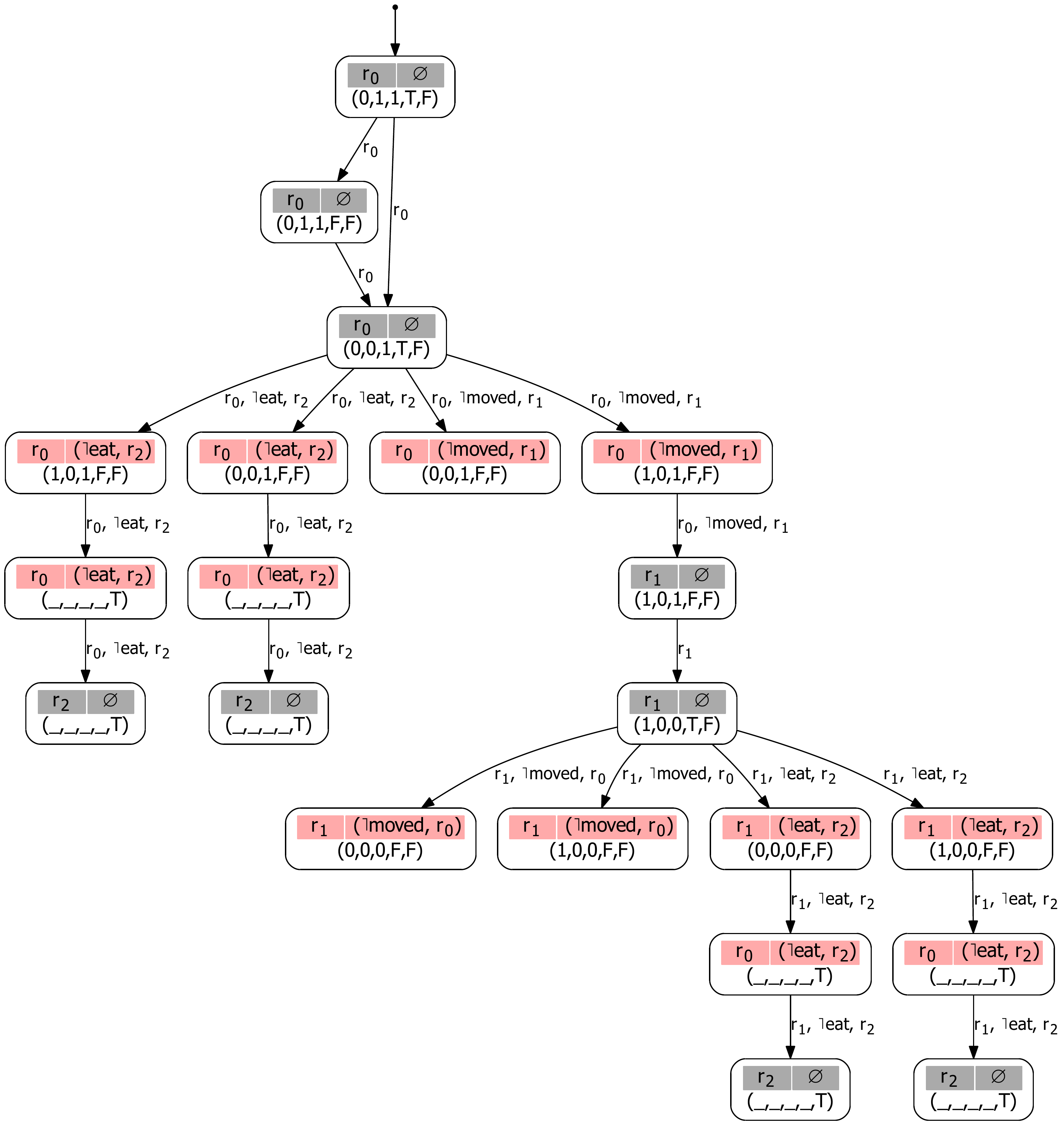}} \hspace*{2cm} \subfloat[]{\includegraphics[scale=0.4]{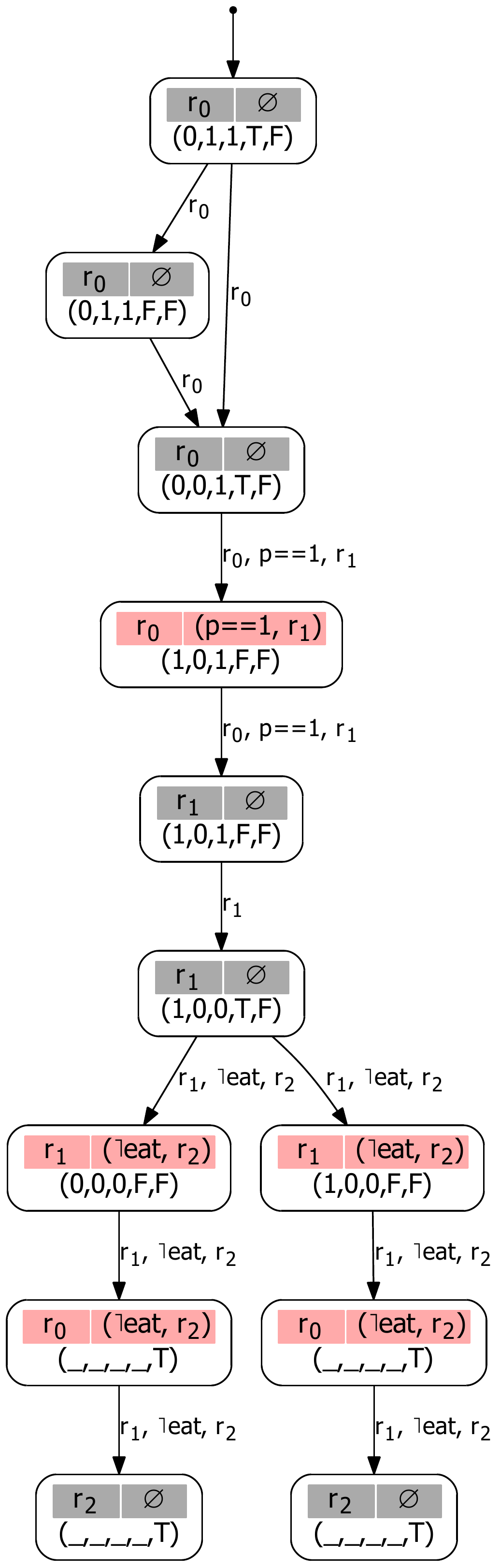}}
\caption{The flat semantics of the two systems $S_0[B]$ (fig.~\ref{fig:sb} (a)) and $S_1[B]$ (fig.~\ref{fig:sb} (b)) in the adaptive 1-predator 2-prey example. Different structural levels lead to different adaptation capabilities. Two adaptation phases (light red marked ones) can be recognized: the first occurs when the predator stops consuming the prey $p_0$, the second when it stops consuming $p_1$. In both systems there always exists an adaptation path leading to a target stable region, but in $S_0[B]$ some paths violate the invariant and cannot proceed. In $S_1[B]$ every adaptation path leads to a target $S$-state.}\label{fig:sb}
\end{sidewaysfigure}

The flattened transitional semantics of the two systems $S_0[B]$ and $S_1[B]$ in the adaptive 1-predator 2-prey food web example presented in Section~\ref{subsect:exmpl} is depicted in Figure~\ref{fig:sb}.
First, we observe that the flat $S_0[B]$ system has a larger state space than the flat $S_1[B]$, due to the higher number of $S$-transitions in $S_0$. In both cases two different adaptation phases can be noticed, the first starting from the flat state $((0,0,1,\mathsf{true},\mathsf{false}),r_0,\emptyset)$ and the second starting from $((1,0,0,\mathsf{true},\mathsf{false}),r_1,\emptyset)$. While in $S_0[B]$ it is possible to adapt to the migration region also in the first phase, in $S_1[B]$ this is possible only in the second phase, i.e.\ when both prey become unavailable.
Moreover in $S_0[B]$, we notice that in each adaptation phase there always exists an adaptation path leading to a target stable region, but some adaptation paths cannot proceed because they violate the invariant. Conversely, in $S_1[B]$ every adaptation path leads to a target $S$-state. Therefore the same behavioural level $B$ possesses different adaptation capabilities, depending on the structure $S$ it is embedded in. These two different kinds of adaptability are formalized in Section~\ref{sect:adaptability}.

Although, depending on the structure $S$, the flat semantics could possibly
lead to a model larger than the behavioural model $B$, the flat $S[B]$-system
lends itself quite naturally to on-the-fly representation techniques. Indeed,
during non-adapting phases it would be necessary to keep in memory just the
subsystem restricted to the set $[[L(r)]] \subseteq B$ of $B$-states that
satisfy the current constraints $L(r)$. On the other hand, as soon as an
adaptation of the form $(q, r, \emptyset) \goes{r,\varphi,r'} (q', r,
\lbrace(\varphi,r')\rbrace)$ takes place, it would be sufficient to store those
$B$-states $q''$ such that $q'' \models \varphi \wedge q'' \not\models L(r')$,
i.e.\ those state where the invariant is met, but the target constraints are
not.

\section{Adaptability relations}\label{sect:adaptability}
The above described transitional semantics for $S[B]$-systems does not guarantee that an adaptation process always leads to a state satisfying the target constraints, or that the system can always start adapting when the current constraints are not met. We characterize this requirements on the adaptability of an $S[B]$-system by means of two binary relations over the set of $B$-states and the set of $S$-states, namely the \textit{weak adaptability relation} $\mathcal{R}_w$ and the \textit{strong adaptability relation} $\mathcal{R}_s$.

Informally, $B$ is \textit{weak adaptable} to $S$ if any active $B$-state $q$ satisfies the constraints imposed by the active $S$-state $r$, or it can start adapting and there exists a finite path reaching a $B$-state $q'$ satisfying the constraints dictated by a target $S$-state $r'$. On the other hand, $B$ is \textit{strong adaptable} to $S$ if any active $B$-state $q$ satisfies the constraints imposed by the active $S$-state $r$, or it can start adapting towards a target $S$-state $r'$ and all paths reach a $B$-state $q'$ satisfying the constraints $L(r')$ in a finite number of transitions.

In the following definitions the notation $\goes{}\!\!^i$ with $i \in \mathbb{N}$ indicates the exponentiation of the transition relation $\goes{}$, i.e.\ $\goes{}\!\!^i = (\goes{})^i = \goes{}(\goes{})^{i-1}$. We use this notation to remark that adaptation paths must be of finite length.

\begin{defn}[Weak adaptability]
\textit{Weak-adaptability} is a binary relation $\mathcal{R}_w \subseteq Q \times R$ defined as follows. Let $q \in Q$ be a $B$-state and $r \in R$ be an $S$-state. Then, $q \ \mathcal{R}_w \ r$ iff
\begin{itemize}
\item $q \models L(r)$ and
\item for all $q' \in Q$, whenever $q \goes{}_B q'$, it holds that either
\begin{itemize}
\item $q' \mathcal{R}_w \ r$, or
\item there exists $q'' \in Q, \varphi \in \Phi(X), r' \in R, i \in \mathbb{N}$, \\$(q, r, \emptyset)\goes{r, \varphi, r'}(q', r, \lbrace (\varphi, r') \rbrace)\goes{r, \varphi, r'}\!\!^i(q'', r', \emptyset)$ and $q'' \mathcal{R}_w \ r'$.
\end{itemize}
\end{itemize}
Let $S[B]$ be an $S[B]$-system. Then $B$ is weak adaptable to $S$ if their
initial states are weak adaptable, i.e.\ $q_0 \ \mathcal{R}_w \ r_0$.
\end{defn}

\begin{defn}[Strong adaptability]
\textit{Strong-adaptability} is a binary relation $\mathcal{R}_s \subseteq Q \times R$ defined as follows. Let $q \in Q$ be a $B$-state and $r \in R$ be an $S$-state. Then, $q \ \mathcal{R}_s \ r$ iff
\begin{itemize}
\item $q \models L(r)$ and
\item for all $q' \in Q$, whenever $q \goes{}_B q'$, it holds that either
\begin{itemize}
\item $q' \mathcal{R}_s \ r$, or
\item $(q, r, \emptyset)\goes{r, \varphi, r'}(q', r, \lbrace (\varphi, r') \rbrace)$ for some $\varphi \in \Phi(X), r' \in R$ \textbf{and} every path starting from \\$(q', r, \lbrace (\varphi, r') \rbrace)$ leads, in a finite number of consecutive $\goes{r, \varphi, r'}$ transitions, to a state $(q'', r', \emptyset)$ such that $q'' \mathcal{R}_s \ r'$.
\end{itemize}
\end{itemize}
Let $S[B]$ be an $S[B]$-system. Then $B$ is strong adaptable to $S$ if their initial states are strong adaptable, $q_0 \ \mathcal{R}_s \ r_0$.
\end{defn}

In the remainder of the paper we will alternatively say that a system $S[B]$ is weak (strong) adaptable, in the sense that $B$ is weak (strong) adaptable to $S$. It is straightforward to see that strong adaptability implies weak adaptability, since the strong version of the relation requires that every adaptation path reaches a target $S$-state, while the weak version just requires that at least one adaptation path reaches a target $S$-state. Now that a relational characterization of adaptability has been given, a concept of equivalence between $B$-states that are adaptable to the same $S$-states naturally arises. Therefore we define the weak adaptation equivalence and the strong adaptation equivalence over the set of $B$-states as follows.

\begin{defn}[Weak adaptation equivalence]
Two $B$-states $q_1, q_2 \in Q$ are said to be equivalent under weak adaptation, written $q_1 \approx_w q_2$, iff for each $S$-state $r \in R$, $q_1 \ \mathcal{R}_w \ r \iff q_2 \ \mathcal{R}_w \ r$.
\end{defn}
\begin{defn}[Strong adaptation equivalence]
Two $B$-states $q_1, q_2 \in Q$ are said to be equivalent under strong adaptation, written $q_1 \approx_s q_2$, iff for each $S$-state $r \in R$, $q_1 \ \mathcal{R}_s \ r \iff q_2 \ \mathcal{R}_w \ r$.
\end{defn}

As discussed in Section~\ref{sect:semantics}, the adaptive 1-predator 2-prey system possesses different adaptation capabilities depending on the structural level $S$. In particular we notice that the system $S_0[B]$ is \textit{weak adaptable}, since in each adaptation phase there always exists an adaptation path leading to a target $S$-state. Nevertheless, it is not strong adaptable because there are adaptation paths that violate the invariant and consequently cannot end adapting. On the other hand, $S_1[B]$ is \textit{strong adaptable}, because every adaptation path leads to a target $S$-state.

\subsection{A logical characterization for adaptability}
In this part we formulate the above introduced adaptability requirements in terms of temporal formulae that can be statically checked on the flat $S[B]$-system.  To this purpose we describe such properties in the well known \textit{CTL (Computational Tree Logic)}~\cite{clarke1986automatic}, a branching-time logic whose semantics is defined in term of states. The set of well-formed CTL formulas are given by the following grammar:
$$\phi ::= \mathsf{false} \ | \ \mathsf{true} \ | \ p \ | \ \neg \phi  \ | \ \phi \wedge \phi \ | \ \phi \vee \phi   \ | \ \mathbf{AX} \phi \ | \ \mathbf{EX} \phi \ | \ \mathbf{AF} \phi \ | \ \mathbf{EF} \phi \ | \ \mathbf{AG} \phi \ | \ \mathbf{EG} \phi \ | \ \mathbf{A}[\phi \mathbf{U} \phi] \ | \ \mathbf{E}[\phi \mathbf{U} \phi],$$
where $p$ is an atomic proposition, logical operators are the usual ones ($\neg, \wedge, \vee$) and temporal operators ($\mathbf{X}$ next, $\mathbf{G}$ globally, $\mathbf{F}$ finally, $\mathbf{U}$ until) are preceded by the universal path quantifier $\mathbf{A}$ or the existential path quantifier $\mathbf{E}$. Starting from a state $s$, CTL operators are interpreted as follows. $\mathbf{AX} \phi$: for all paths, $\phi$ holds in the next state; $\mathbf{EX} \phi$: there exists a path s.t. $\phi$ holds in the next state; $\mathbf{AF} \phi$: for all paths, $\phi$ eventually holds; $\mathbf{EF} \phi$: there exists a path s.t. $\phi$ eventually holds; $\mathbf{AG} \phi$: for all paths, $\phi$ always holds; $\mathbf{EG} \phi$: there exists a path s.t. $\phi$ always holds; $\mathbf{A}[\phi_1 \mathbf{U} \phi_2]$: for all paths, $\phi_1$ holds until $\phi_2$ holds; and $\mathbf{E}[\phi_1 \mathbf{U} \phi_2]$: there exists a path s.t. $\phi_1$ holds until $\phi_2$ holds).

In the following we provide the CTL formulas characterizing a weak adaptable and a strong adaptable $S[B]$-system. Formulas are evaluated over the flat semantics and we employ two atomic propositions: $adapting$, to denote an adapting state, and $steady$ to denote a steady one. More formally, we define, given a flat $S[B]$ system $F$ and a state $s=(q_s,r_s,\rho_s)$ of $F$, 
$$
\langle F, s \rangle \models_{\mathrm{CTL}} adapting \iff (q_s,r_s,\rho_s) \goes{r_{s}, \varphi, r'}
$$ 
for some $\varphi \in \Phi(X)$ and $r' \in R$; moreover, 
$$
\langle F, s \rangle \models_{\mathrm{CTL}} steady \iff (\rho_{s} = \emptyset \wedge (q_s,r_s,\rho_s) \;\;\;\;\;\;\;  \not \!\!\!\!\!\!\!\!\!\! \goes{r_{s}, \varphi, r'}).
$$ 
Additionally, the connective $\phi_1 \implies \phi_2$ has the usual meaning: $\neg \phi_1 \vee \phi_2$.
\begin{itemize}
\item \textbf{Weak adaptability:} there is a path in which, as soon as adaptation starts, there exists at least one path for which the system eventually ends the adaptation phase leading to a target $S$ state.
\begin{equation}\label{eq:weak}
\mathbf{EG}(adapting \implies \mathbf{EF} \ steady)
\end{equation}
\item \textbf{Strong adaptability:} for all paths, it always holds that whenever the system is in an adapting state, for all paths it eventually ends the adaptation phase leading to a target $S$ state.
\begin{equation}\label{eq:strong}\mathbf{AG}( adapting \implies \mathbf{AF} \ steady)\end{equation}
\end{itemize}
\begin{prop}[Equivalent formulations of weak adaptability]
Let $S[B]$ be an $S[B]$-system. Then, $S[B]$ is weak adaptable if and only if $S[B]$ satisfies the weak adaptability CTL formula (equation~\ref{eq:weak}). Formally, $q_0 \ \mathcal{R}_s \ r_0 \iff \langle F, f_0 \rangle \models_{CTL} \mathbf{EG}(adapting \implies \mathbf{EF} \ steady)$, where $F$ is the flat semantics of $S[B]$, $q_0$, $r_0$ and $f_0$ are the initial states of the behavioural level $B$,  of the structural level $S$ and of the flattened system $F$, respectively.
\end{prop}
\begin{prop}[Equivalent formulations of strong adaptability]
Let $S[B]$ be an $S[B]$-system. Then, $S[B]$ is strong adaptable if and only if $S[B]$ satisfies the strong adaptability CTL formula (equation~\ref{eq:strong}). Formally, $q_0 \ \mathcal{R}_w \ r_0 \iff \langle F, f_0 \rangle\models_{CTL} \mathbf{AG}( adapting \implies \mathbf{AF} \ steady)$, where $F$ is the flat semantics of $S[B]$, $q_0$, $r_0$ and $f_0$ are the initial states of the behavioural level $B$, of the structural level $S$ and of the flattened system $F$, respectively.
\end{prop}

Note that since we assume that the behavioural and the structural state machines are finite state, then the CTL adaptability properties can be model checked. This means that the defined notions of weak and strong adaptability are decidable.

\section{Discussion and conclusion}\label{sect:conclusion}
In this work we presented $S[B]$-systems, a general multi-level model for
self-adaptive systems, where the lower $B$-level is a state machine describing
the behaviour of the system and the upper $S$-level is a second-order state
machine accounting for the dynamical constraints with which the system has to
comply. Higher-order $S$-states identify stable regions that the $B$-level may
reach by performing adaptation paths. An intriguing (but here simplified) case
study from ecology has been provided to demonstrate the capabilities of
$S[B]$-systems: the adaptive 1-predator 2-prey system. The semantics of the
multi-level system is given by a flattened transition system and two different
concepts of adaptability (namely, weak and strong adaptability) have been
formalized, both in a relational flavour and with CTL formulas that can be model
checked. We report that this work gives a formal computational characterization
of self-adaptive systems, based on concepts like multiple levels and
higher-order structures that are well-established in the science of complex
systems.

Note also that in this work we defined in details just two levels,
namely the $S$-level and the $B$-level. However, our approach can be easily
extended in order to consider multiple levels arising from the composition of
multiple $S[B]$-systems. Let $\lbrace S^n[B^n]_i \ | \ i\in I\rbrace$ be a set
of $S[B]$-systems at a certain level $n$. Their parallel composition would be
defined as $\|_{i \in I} \ S^n[B^n]_i$. Then, if we let $B^{n+1} = \|_{i \in I}
\ S^n[B^n]_i$ be the behavioural state machine at level $n+1$, an higher-level
$S[B]$-system $S^{n+1}[B^{n+1}]$ can be built by defining a structure $S^{n+1}$
at level $n+1$, together with a set of observable variables $X^{n+1}$ and with
an observation function $\mathcal{O}^{n+1}$.

The present work is just an initial attempt and several extensions can be
integrated into the model in the next future. First, the definition of a
higher-level algebraic language for specifying $S[B]$-systems would be useful in
order to handle more complex and larger models of adaptive systems.
Additionally, we are currently investigating further adaptability relations and
different models for the structural level, where adaptation can occur not only
when no possible future behaviours satisfy the current constraints, but also
when stability conditions are met. Then, another possible research direction
would be embedding quantitative aspects into the two levels of an $S[B]$-system.
In this way, an $S$-transition would have associated a measure of its
cost/propensity, for distinguishing the adaptation paths more likely to occur
(e.g. in the 1-predator 2-prey example, the predator adapting its diet), to
those less probable (e.g. the predator migrating even under prey availability
conditions).

Finally we assume that the reciprocal knowledge between the two levels is
limited: they see each other as black-box systems. However, this approach could
be extended in order that the structure $S$ has a more comprehensive knowledge
of the behaviour $B$. Under the \textit{white-box assumption}, the structure
could act as a sort of monitor that is able to statically check the behavioural
model for properties of safe adaptation. In this way, the system will know in
advance if an adaptation path eventually leads to a target $S$-state and if not,
it will avoid that path. In other words, runtime model checking techniques
allows the system to behave in an \textit{anticipatory way}. Anticipation is a
crucial property in complex self-adaptive systems, since it makes possible to
adjust present behaviour in order to address future faults. A well-know
definition is given by Rosen~\cite{rosen1985anticipatory}: \textit{``An
anticipatory system is a system containing a
predictive model of itself and/or its environment, which allows it to change
state at an instant in accord
with the model's predictions pertaining to a later instant''}. In the settings
of $S[B]$-systems, the predictive model of the system could be the behavioural
level itself, or a part of it if we assume that $S$ does not have a complete
knowledge of $B$ and is able to ``look ahead'' only at a limited number of
future steps. The verdict of runtime model checking would be what Rosen refers
to as model's predictions.

\subsection*{Acknowledgements}
This work was partially supported by the project  ``TOPDRIM: \textit{Topology Driven Methods for Complex Systems}'' funded by the European Commission (FP7 ICT FET Proactive - Grant Agreement N.\ 318121). The authors thank Marianna Taffi for helping in the definition of the case study.
\bibliographystyle{eptcs}
\bibliography{self}
\end{document}